\documentclass[journal=jacsat,manuscript=article]{achemso}

\usepackage{chemformula} 
\usepackage[T1]{fontenc} 
\usepackage[version=3]{mhchem} 
\usepackage{amssymb,stackengine,graphicx}
\usepackage{siunitx}
\usepackage[numbers]{natbib}

\author{Vincenzo Calabrese}
\email{vincenzo.calabrese@oist.jp}
\author{Simon J. Haward}
\author{Amy Q. Shen}
\email{amy.shen@oist.jp}
\affiliation[Unknown University]
{Okinawa Institute of Science and Technology, Onna-son, Okinawa 904-0495, Japan}

\title[An \textsf{achemso} demo]
  {Decoupling the effects of shear and extensional flows on the alignment of colloidal rods  }

\begin{document}

%
%

\begin{abstract}
Cellulose nanocrystals (CNC) can be considered as model colloidal rods and have practical applications in the formation of soft materials with tailored anisotropy. Here, we employ two contrasting microfluidic devices to quantitatively elucidate the role of shearing and extensional flows on the alignment of a dilute CNC dispersion. Characterization of the flow field by micro-particle image velocimetry is coupled to flow-induced birefringence analysis to quantify the deformation rate--alignment relationship. The deformation rate required for CNC alignment is 4$\times$ smaller in extension than in shear. Alignment in extension is independent of the deformation rate magnitude, but is either 0$^\circ$ or 90$^\circ$ to the flow, depending on its sign. In shear flow the colloidal rods orientate progressively towards 0$^\circ$ as the deformation rate magnitude increases. Our results decouple the effects of shearing and extensional kinematics at aligning colloidal rods, establishing coherent guidelines for the manufacture of structured soft materials. 
\end{abstract}

\section{INTRODUCTION}
Colloidal rods have received long-lasting attention as building-blocks for complex materials because of their effective gelling properties and their ability to orient upon hydrodynamic forces, allowing the manufacture of materials with tailored anisotropy.\citep{Solomon2010,Rosen2020b,Calabrese2020b} Soft materials with anisotropic orientation have shown a large number of advantages, such as their mechanical strength,\cite{Hakansson2014,Nechyporchuk2019} structural color,\citep{Liu2014} electrical and thermal conductivity,\citep{Kiriya2012,Xin2019} and directional control of cell growth.\citep{DeFrance2017}

To achieve directionality in soft materials, microfluidic platforms have been extensively used to control hydrodynamic forces, mainly through shearing and extensional-dominated flows, to aid particle alignment while keeping negligible inertia effects. The relative strength between diffusion and hydrodynamic forces is typically described by the P\'{e}clet number $Pe=|E|/Dr$, where $|E|$ is the characteristic deformation rate, and the rotational diffusion coefficient for non-interacting rods, $Dr$, is described as\citep{LangStiffness}
\begin{equation}
   Dr=\frac{3 k_b T \mathrm{ln}(l/d_{eff})}{\pi \eta_{s} l^3}~,
   \label{eqn:Dr}
\end{equation}
where $k_b$ is the Boltzmann constant, $T$ the temperature, $\eta_{s}$ the solvent shear viscosity, $l$ is the length and $d_{eff}$ the effective diameter of the rod which accounts for the thickness of the electric double layer. For $Pe<1$ the particles are dominated by Brownian fluctuations, whilst for $Pe>1$, the particles are perturbed by the flow field. It has been shown, experimentally and theoretically, that shearing flows enable a gradual alignment of anisotropic particles towards the flow direction, where particles align with a preferential angle of 45$^\circ$ at $Pe\simeq1$ and achieve orientation parallel to the flow direction, 0$^\circ$, at $Pe\gg1$.\citep{Vermant2001,Dhont2007,Reddy2018,Winkler2004} Although shear-dominated flows are relatively simple to study \textit{via} conventional rheo-optics techniques, ``pure'' extensional flows in conditions where shear-forces are negligible are experimentally difficult to assess.\citep{Haward2016,Lang2019a} Consequently, much less is known on how extensional rates affect the orientation of particles in shear-free conditions. 
Nonetheless, it has been shown that extensional rates enable an additional control on particle orientations, inducing, for instance, particle alignment perpendicular to the flow direction, which is not possible in shearing flows.\citep{Trebbin2013,Kiriya2012,Qazi2011,Pignon2003} Of considerable importance, is the work of Corona et al., who investigated the effects of shearing and extension-dominated flows on the alignment of a concentrated suspension of colloidal rods using a fluidic four-roll mill device.\citep{Corona2018} However, no significant differences could be discerned between shearing and extensional-dominated flows as excluded volume effects were likely dominant at the high concentration tested. Recent work from Ros\'{e}n et al.\citep{Rosen2020a} elucidated that for extension-dominated flows, generated in flow-focusing and converging channels, a reduced rotational diffusion coefficient was obtained when compared to shearing flows. Nonetheless, difficulties rising from interparticle interactions, particle flexibility, and a non uniform extensional rate within the microfluidic channel, hindered a comprehensive quantitative examination to decouple the effects of shearing and extensional dominated flows. To date, a quantitative comparison between shear and extensional rate--driven alignment of rod-like particles, in conditions where interparticle interactions are negligible, is still missing. 

In this article we couple quantitative flow field measurements with state-of-the-art flow-induced birefringence analysis to elucidate the role of shear and extensional rates at aligning rod-like particles. A dilute cellulose nanocrystal (CNC) suspension consisting of negatively charged rigid rod-like nanoparticles is used as a model system.\citep{Hasegawa2020a}

\section{MATERIALS AND METHODS}
\subsection{Test fluid}
The CNC was purchased from CelluForce (Montreal, Canada) as an aqueous 5.6 wt\% stock dispersion at pH 6.3. A 0.1wt\% CNC dispersion was prepared by dilution of the stock dispersion with deionised water and used without further treatment. Extensive characterization of CNC from the same industrial producer is described by Bertsch et al.\citep{Bertsch2017,Bertsch2019} and Reid et al.\citep{Reid2017}
\subsection{Shear Rheometry}
Steady shear rheology of the 0.1 wt\% CNC dispersion was measured using a strain-controlled ARES-G2 rotational shear rheometer (TA Instruments Inc.) equipped with a double gap geometry (with an inner and outer gap of 0.81 and 1.00~mm, respectively) composed of a stainless steel bob and a hard-anodized aluminum cup. The dispersion was covered with a solvent trap and measurements were performed at 25$^\circ$C (controlled by an advanced Peltier system with temperature accuracy of $\pm$~0.1$^\circ$C).
A microfluidic slit rheometer (m-VROC RheoSense Inc.), equipped with an A10 pressure cell, 3 mm wide and \SI{100}{\micro\meter} high, was used to access the rheological response of the test fluid at high values of shear rate (up to $\dot\gamma\approx 3 \times 10^4$ s$^{-1}$). The experiment was carried out at $25^{\circ}$C (controlled via an external circulating water bath with temperature accuracy of $\pm$~0.1$^\circ$C).

\subsection{Atomic force microscopy (AFM)}
A 0.01~wt\% CNC dispersion was drop-casted on a mica substrate and imaged using an atomic force microscope (Dimension ICON3, Bruker) in tapping mode. The distribution of the particle contour length, $l$, and diameter, $d$, was extrapolated by tracking 976 isolated particles using an open-source code,  FiberApp.\citep{Usov2015a} The value of $d$ was obtained from the AFM height profile.  

\subsection{Microfluidic platforms}
\begin{figure}[h!]
\includegraphics[width=8cm]{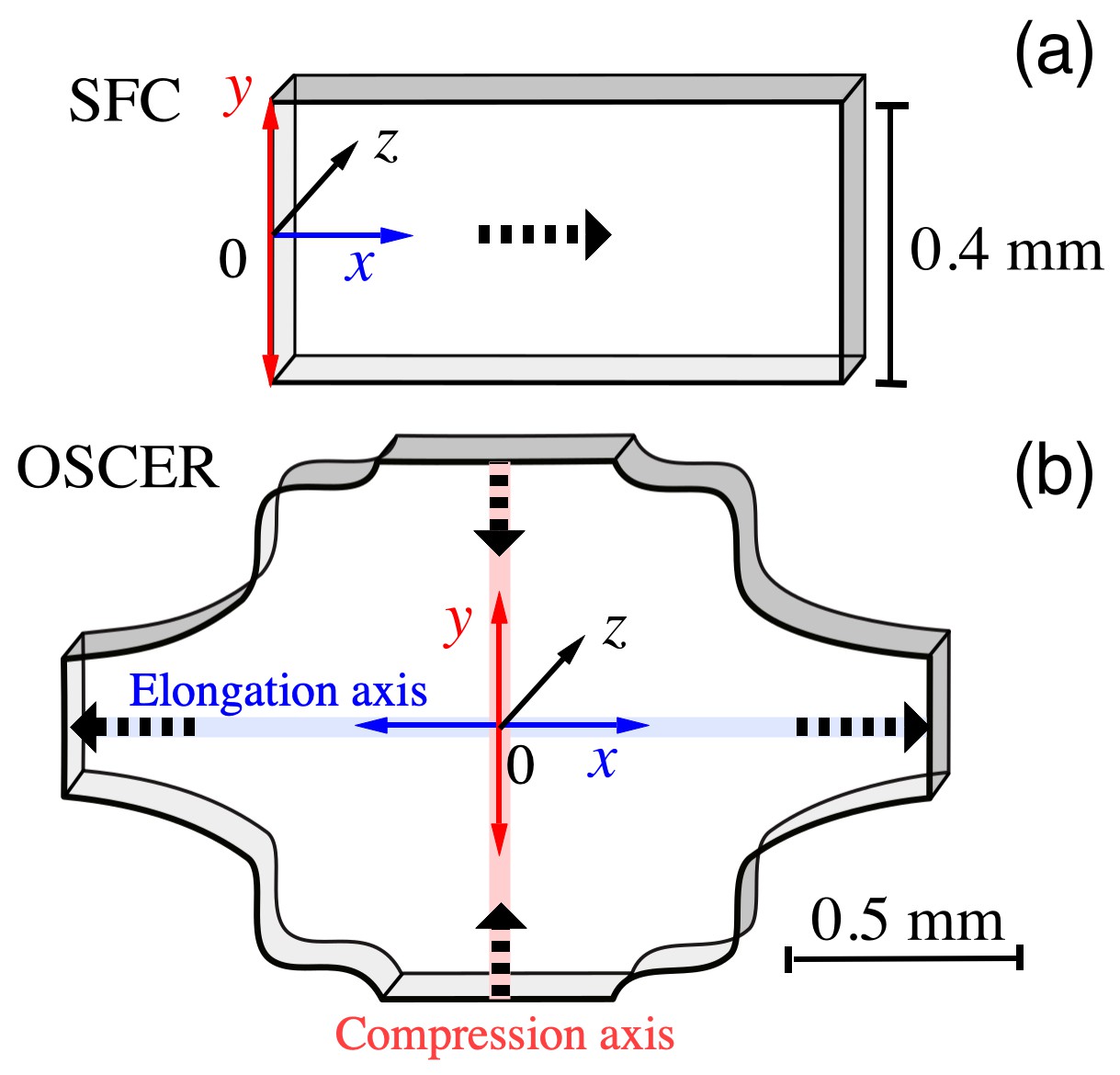}
\caption{Schematic diagrams of (a) the shearing flow-dominated channel (SFC) and (b) the optimized shape cross-slot extensional rheometer (OSCER) with respective coordinate system and scale bar. The dashed arrows indicate the flow direction.}
  \label{fgr:Channels}
\end{figure}

A shearing flow-dominated channel (SFC) and an optimized shape cross-slot extensional rheometer (OSCER)\citep{Haward2012a,Haward2016} were used to generate two-dimensional (2D) flows that, in discrete areas of the geometries, provide good approximations to purely shearing and purely extensional flows, respectively (Figure~\ref{fgr:Channels}).
The SFC consists of a straight channel of fused silica glass with a rectangular cross section (length $L=25$ mm along the $x$-axis, height $H=2$~mm along the $z$-axis, and width $W=0.4$~mm along the $y$-axis, resulting in an aspect ratio $\alpha=H/W=5$), fabricated with a selective laser-induced etching (SLE) technique\citep{Burshtein2019,Haward2019} (Figure~\ref{fgr:Channels}a). The OSCER is based on a planar cross-slot geometry with two incoming and outgoing flows placed orthogonal to each other, as described by Haward et al.\citep{Haward2012a,Haward2016} and depicted in Figure~\ref{fgr:Channels}b. The device has a height of $H=2.1$~mm along the $z$-axis and a channel width of $W=0.2$ mm at the inlets and outlets, yielding $\alpha=10.5$, generating a good approximation of a 2D flow field which is extensional-dominated in a large region around the stagnation point, at $x=y=0$ (see coordinate system in Figure~\ref{fgr:Channels}b). The fluid elements are compressed along the $y$-axis and extended along the $x$-axis (referred to as the compression and elongation axes, respectively, Figure~\ref{fgr:Channels}b). 

The flow inside the channels is driven by Nemesys low-pressure syringe pumps (Cetoni, GmbH) and Hamilton Gastight syringes, which infuse the liquid at the inlet and withdraw it at an equal and opposite volumetric rate, $Q$ (m$^3$~s$^{-1}$), from the outlet. To ensure steady flows, the highest flow rates were set at  Reynolds numbers, $Re=\rho U W/\eta$, of 15.4 and 6.4, for the SFC and OSCER, respectively, considering a fluid density, $\rho=1000$ kg~m$^{-3}$, an average flow velocity as $U=Q/HW$ (m~s$^{-1}$) and a fluid viscosity $\eta=1.3$ mPa~s as determined by rheological measurements (detailed in the following section). The flows were equilibrated for at least 5~s before all measurements and confirmed as being steady by inspection of the micro-particle image velocimetry ($\mu$-PIV, see next section) flow fields prior to their time-averaging. All the experiments were carried out at the ambient laboratory temperature ($ 25 \pm 1^{\circ}$C).

\subsection{Micro-particle image Velocimetry ($\mu$-PIV)}
The flow fields in the SFC and OSCER devices were obtained using time averaged $\mu$-PIV of the test fluid seeded with \SI{1.1}{\micro\meter} fluorescent particles (Fluoro-Max\textsuperscript{TM}, Thermo Fisher), to a concentration of $\approx0.02$ wt\%. The $\mu$-PIV measurements were conducted using a volume illumination system (TSI Inc., MN) installed on an inverted microscope (Nikon Eclipse Ti). Nikon PlanFluor objective lenses of 4$\times$ and 10$\times$ with numerical apertures of NA=0.13 and 0.30 were used for the OSCER and SFC device, respectively. Each geometry was placed with $z$-axis parallel to the light source and for all the flow rates tested, a sequence of at least 100 image pairs were acquired at the midplane of the geometries ($z=1$ mm). For the SFC, images were acquired at a distance of $\approx L/2$ from the inlets to ensure fully developed velocity profiles. The average displacement of the seeded particles between the two images in each pair was kept constant at $\approx$ 4 pixels. The measurement depths, $\delta z_m$, corresponding to the depth over which the seeded particles contribute to the determination of the velocity field was $\delta z_m =150~ $\SI{}{\micro\meter} and \SI{35}{\micro\meter}, for the OSCER and SFC, respectively.\citep{Meinhart2000} Cross-correlation between image pairs provided velocity vectors on a square grid with a spatial resolution of 2.0 and 0.8 \SI{}{\micro\meter}/pixel for the OSCER and SFC, respectively. Data analysis was performed using a custom-made Matlab routine. 

\subsection{Flow-induced birefringence (FIB)}
Flow-induced birefringence (FIB) measurements were performed using an Exicor MicroImager (Hinds Instruments, Inc., OR). Monochromatic light of wavelength $\lambda=450$~nm was shone through a linear polarizer at 0$^\circ$, a photoelastic modulator (PEM) at 45$^\circ$, the SFC or the OSCER containing the testing fluid, a PEM at 0$^\circ$ and a linear polarizer at 45$^\circ$, in the order given. The geometries were imaged using a 5$\times$ objective in the same position as described for $\mu$-PIV. The instrument performs Mueller matrix decomposition, determining the elements of 4$\times$4 Mueller matrices using a stroboscopic light source. The retardance $R$, describing the total phase shift occurring between the two orthogonally polarized light beams, and the orientation of the slow optical axis (extraordinary ray), $\theta$, were obtained from a total of 7 images acquired at 1 frame/s. The retardance, $R$ (measured in nm), was then converted to birefringence as $\Delta n =R/H$. The background value of $\Delta n$ was determined for the test fluid at rest and subtracted for all the analysis presented. The background value determined in both geometries was $\Delta n \approx 6\times10^{-7}$ and comparable to the instrument resolution of $\Delta n \approx 3\times10^{-7}$. The spatial resolution of the measurement was $\approx 2$ \SI{}{\micro\meter}/pixel and data analysis was performed using a custom-made Matlab routine.

\section{RESULTS AND DISCUSSION}
\subsection{Characterization of the test fluid}
\begin{figure}[h!]
\includegraphics[width=8cm]{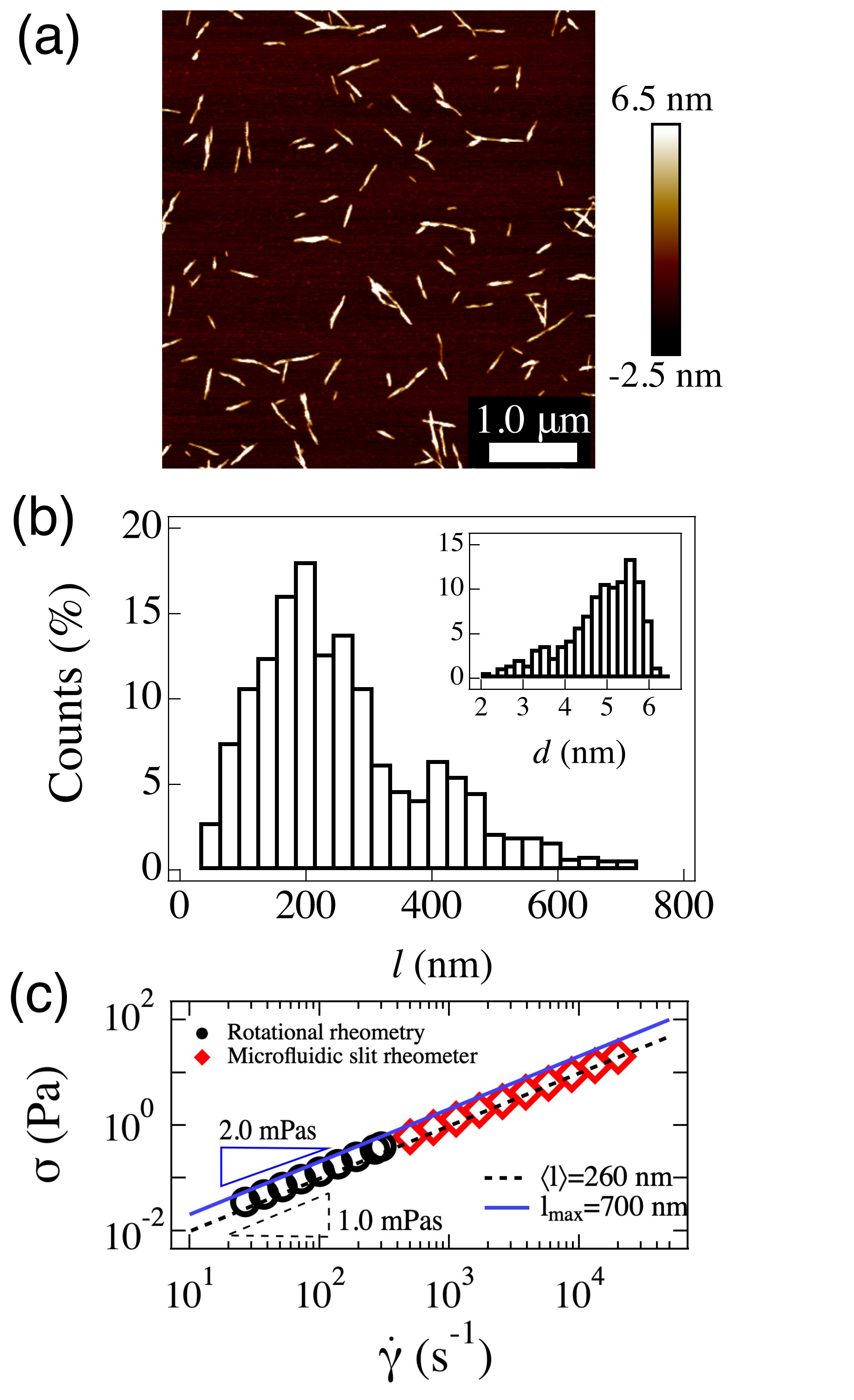}
\caption{(a) Tapping-mode atomic force microscopy (AFM) image of a drop-casted 0.01wt\% CNC dispersion on mica substrate. (b) Particle contour length, $l$, and particle diameter, $d$ (inset in (b)), distributions as obtained from the AFM counting of 976 isolated particles. (c) Flow curve of the 0.1 wt\% CNC dispersion presented as shear stress, $\sigma$ \textit{vs} shear rate, $\dot\gamma$. The black circles are data obtained using a rotational strain controlled rheometer (ARES-G2). The data represented by the diamonds are obtained using a microfluidic slit rheometer (m-VROC). The dashed (black) and solid (blue) lines are the predictions of the zero shear viscosity from eqn.\ref{eqn:eta} using values of average length, $\langle l \rangle$, and maximum length, $l_{max}$, respectively, as obtained from AFM analysis.  }
  \label{fgr:Fig1}
\end{figure}

We begin by estimating the volume density, $\nu=N / V$, of the 0.1 wt\% CNC dispersion, where $N$ is the number of rods and $V$ is the sample volume. In suspensions of monodisperse rod-like particles, the volume density, $\nu$, is commonly used to distinguish between the dilute, semi-dilute and concentrated regimes.\citep{Solomon2010} For monodisperse rods, the calculation of $\nu$ is straightforward as a single particle length ($l$) and diameter ($d$) are sufficient to describe the whole particle population. Contrarily, for a polydisperse distribution of rods such as CNC, the large span of lengths must be considered for a more accurate estimation of $\nu$. Thus, the size distribution of the CNC was extrapolated from atomic force microscopy (AFM) images (Figure~\ref{fgr:Fig1}a,b). The average contour length was $\langle l \rangle = 260 \pm180$ nm and the average diameter $\langle d \rangle = 4.8 \pm1.8$ nm. The effective volume density, $\nu_e$, was thereafter estimated as
\begin{equation}
   \nu_e=\frac{N_e}{V}=\frac{\sum_{i=1}^{l_{max}} \bigg(\frac{V_{CNC}\phi_i}{Vcyl_i}\bigg)}{V},
   \label{eqn:nueff}
\end{equation}
where $N_e$ is the effective number of rods, $l_{max}$ is the longest detected contour length (700~nm), $V_{CNC}$ is the volume of the CNC in the sample (which can be estimated using a density of 1500~kg~m$^{-3}$),\citep{RyanWagnerArvindRaman2010} $\phi_i$ is the volume fraction of the rods with length $i$ and $Vcyl_i$ is the volume occupied by a single rod with length $i$, for which a cylindrical morphology can be approximated. It is noted that the distribution of the CNC diameters, $d$, is not accounted for in eqn.~\ref{eqn:nueff} since it has only a minimal effect on the estimation of $\nu_e$ of slender objects and $\langle d \rangle$ is used to obtain $Vcyl_i$. 
For the 0.1 wt\% CNC, the $\nu_e\approx 1/ \langle l \rangle^3$, indicating that at this concentration, the CNC dispersion is at the onset of the semi-dilute regime, where the particles rarely interact (although more frequently than in the dilute regime).\citep{LangStiffness,Lang2019} The absence of pronounced interparticle interactions was reported by Bertsch et al. for concentrations below 0.5~wt\%, as shown by small angle X-ray scattering studies of CNC from the same source as that used in the present work.\citep{Bertsch2019} In addition, from the AFM images, a persistence length, $l_p\approx30\langle l\rangle$ was extrapolated using the method of the mean-squared midpoint displacement (MSMD) within the FiberApp routine and detailed by Usov and Mezzenga,\citep{Usov2015a} indicating that CNC can be well described as rigid rods. A value of $l_p\gg l$ was also obtained when $l_p$ was calculated as\citep{Usov2015}
\begin{equation}
   l_p=\frac{\pi \langle d \rangle^4 G}{64 k_b T},
   \label{eqn:lp}
\end{equation}
using the reported values of the CNC elastic modulus, $G$, between 5 and 150 GPa.\citep{RyanWagnerArvindRaman2010}
The steady shear rheology of the 0.1~wt\% CNC dispersion is shown in Figure~\ref{fgr:Fig1}c. The shear stress, $\sigma$, of the 0.1 wt\% CNC suspension followed a linear relationship with the shear rate, $\dot\gamma$, as for a Newtonian fluid with shear viscosity $\eta= 1.3$~mPa~s. Anisotropic particles are expected to exhibit a viscosity plateau at low shear rate, namely the zero shear viscosity, $\eta_0$, followed by a shear thinning behavior up to a second viscosity plateau at higher shear rates, $\eta_\infty$.\citep{Lang2019,Lang2020b,Tanaka2014,Kobayashi2011,Lang2016} From eqn.~\ref{eqn:Dr} it is clear that longer rods will align at lower shear rates than shorter rods since $Dr\propto 1 / l^3$. As such, in polydisperse suspensions of non-interacting colloidal rods, the onset of shear thinning is dictated by the longest population of rods. 
Substituting $l$ with $l_{max}$ and accounting for the contribution of the electric double layer ($\delta{d}=22.6$~nm)\citep{Bertsch2019} to the effective diameter ($d_{eff}\approx~\delta{d} +\langle d \rangle$) in eqn.~\ref{eqn:Dr}, we estimated the expected onset of the shear thinning at $\dot\gamma\simeq 40 \ $~s$^{-1}$ ($Pe~\simeq ~1$). 
However, at this CNC concentration, the value of viscosity of the test fluid is close to that of the solvent shear viscosity, $\eta_{s}$, making practically impossible to capture the shear thinning region expected for anisotropic particles. For a dilute dispersion of monodisperse rods, $\eta_0$ can be estimated as\citep{LangStiffness}
\begin{equation}
  \eta_0\simeq \eta_{s}+\nu_e k_b T \bigg(\frac{1}{30} \frac{1}{Dr} \bigg).
   \label{eqn:eta}
\end{equation}
When the value of $l$ in eqn.~\ref{eqn:Dr} is substituted with either $\langle l \rangle$ or $l_{max}$, the predicted values of zero shear viscosity well encompass the experimental data (Figure~ \ref{fgr:Fig1}c). The lack of any further contribution needed to account interparticle interactions in eqn. \ref{eqn:eta} suggests that the 0.1~wt\% CNC dispersion can be considered in the dilute regime.  

\subsection{Flow profiles}
The bulk rheometry gives a good indication that the diluted CNC dispersion behaves as a Newtonian fluid. Consequently, no information regarding the onset of alignment, as usually associated with the onset of shear thinning, can be discerned from the flow curve. To define a deformation rate--alignment relationship of the dilute CNC dispersion we focus on the control of the shear rate, $\dot\gamma$, and extensional rate, $\dot\varepsilon$ in two separate microchannels. 
\begin{figure}[h!]
\includegraphics[width=16cm]{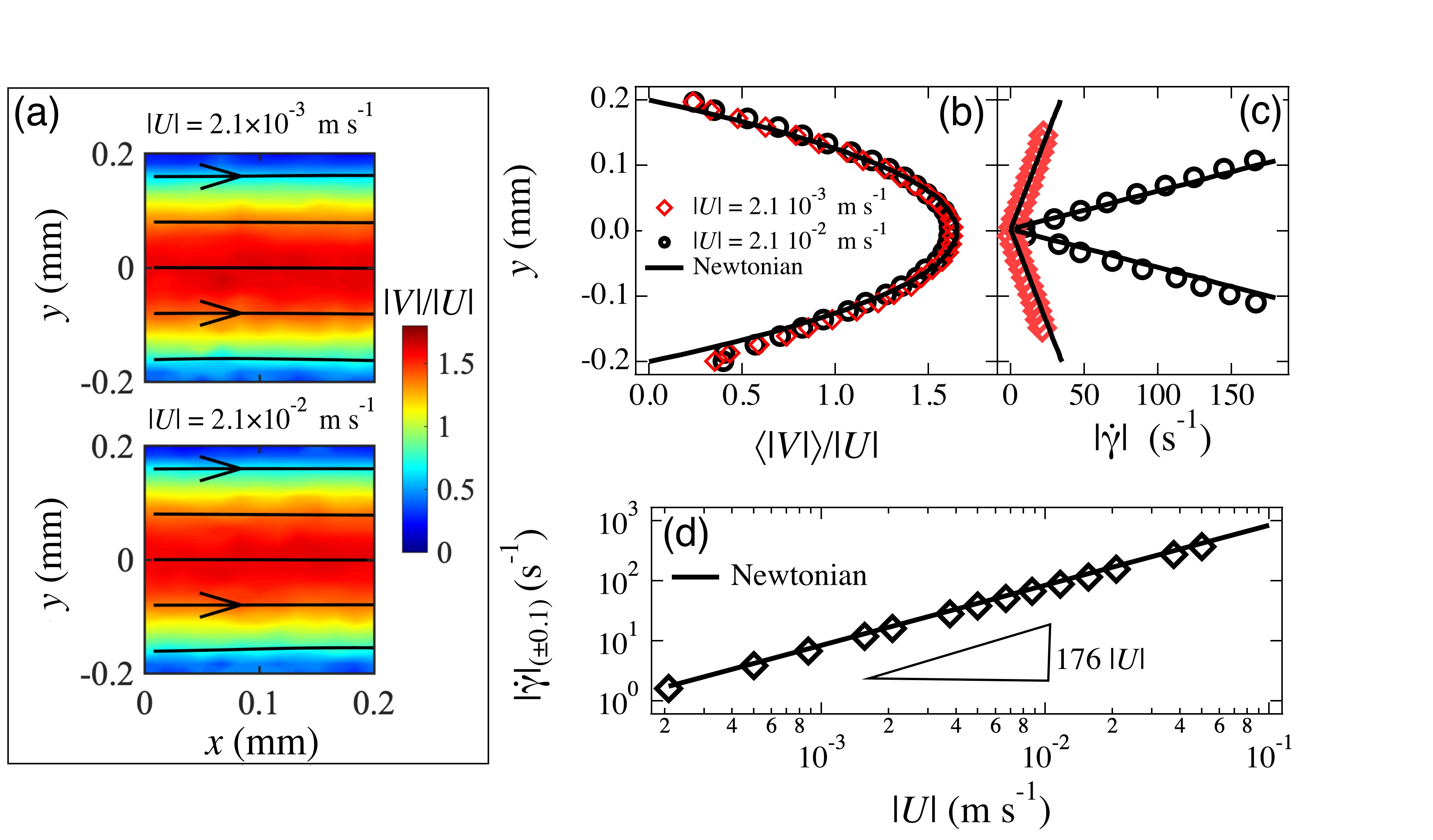}
\caption{(a) Time averaged results of flow velocimetry ($\mu$-PIV) with superimposed streamlines for the SFC containing the 0.1~wt\% CNC dispersion. Two representative average flow velocities, $|U|$, are displayed. (b) Normalized spatially averaged velocity profiles taken across the $y$-axis of the channel, for values of $|U|$ as in (a). (c) Magnitude of shear rate, $|\dot\gamma|=\partial|V_x|/\partial|y|$, as obtained from the velocity profile displayed in (b). (d) Magnitude of the shear rate obtained from the average of $|\dot\gamma|$ at $y=\pm0.1$ mm, $|\dot\gamma|_{(\pm0.1)}$, as a function of the average flow velocity. In the panels (b), (c), and (d), the solid lines are the infinite series analytical solution for creeping Newtonian flow.\citep{Shah}}
  \label{fgr:Fig2}
\end{figure}
The SFC was used to provide a well approximated 2D  shearing flow. Figure~\ref{fgr:Fig2}a shows two representative time averaged flow fields obtained by $\mu$-PIV at the midplane ($z=1$~mm) of the SFC. The velocity magnitude $|V|$, as measured by $\mu$-PIV, is scaled by the average flow velocity $|U|$. The flow field displays a velocity gradient across the channel width ($y$-axis) with the greater velocity along the centerline. This is evident when plotting the spatially averaged velocity, namely $\langle|V|\rangle$ (determined by averaging $|V|$ along 0.2 mm of the $x$-axis) as $\langle|V|\rangle/|U|$ \textit{vs} the channel width ($y$-axis), where a parabolic (Poiseuille) flow profile is displayed (Figure~\ref{fgr:Fig2}b). 
The measured flow profile is in quantitative agreement with an infinite series analytical solution for creeping Newtonian flow\citep{Shah} as depicted by the solid line in Figure~\ref{fgr:Fig2}, consistent with the Newtonian-like behavior described by the rheological measurement in Figure~\ref{fgr:Fig1}. From the velocity profile displayed in Figure~\ref{fgr:Fig2}b, we compute the shear rate profile as $|\dot\gamma|=\partial|V_x|/\partial|y|$, where $|V_x|$ is the $x$-component of velocity (Figure~\ref{fgr:Fig2}c). The magnitude of the shear rate is also in good agreement with the expectation for a Newtonian fluid (black lines). Since the shear rate varies substantially along the channel $y$-axis, we select the value of the shear rate at the location $y=\pm$~$0.1$~mm, namely $|\dot\gamma|_{(\pm 0.1)}$ (computed as the average of the $|\dot\gamma|$ at $y=0.1$ mm and $y=-0.1$~mm), which is the mid-point in $|y|$ between the minimum and the maximum value of $|\dot\gamma|$. The location $y=\pm$~$0.1$~mm is also far from the channel side walls, where the $\mu$-PIV limitations become evident, while still providing relatively high values of~shear~rate. 
\begin{figure}[ht!]
\includegraphics[width=16.5cm]{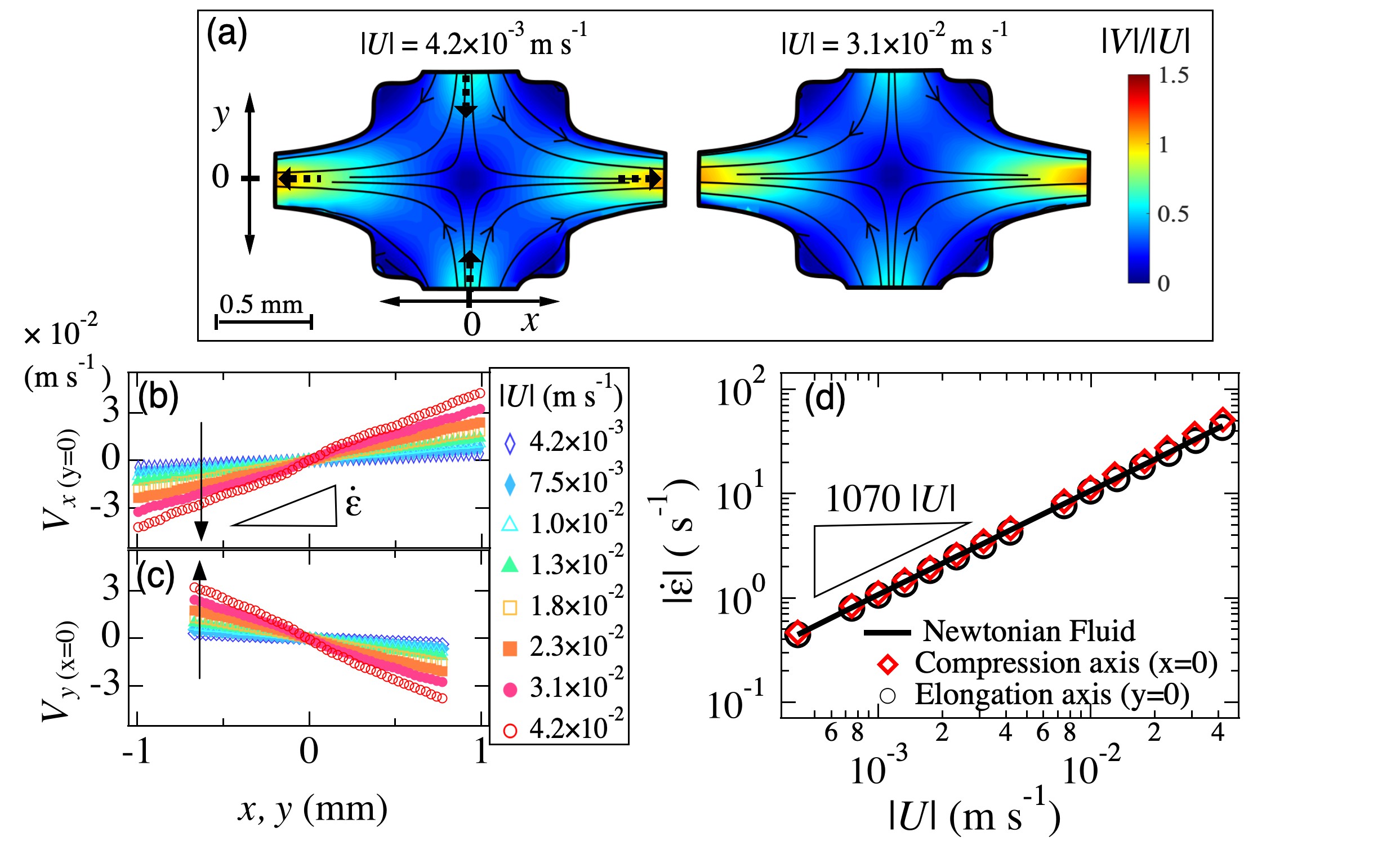}
\caption{(a) Time averaged results of flow velocimetry ($\mu$-PIV) with superimposed streamlines for the OSCER containing the 0.1~wt\% CNC dispersion. Two representative average flow velocities, $|U|$, are displayed. (b) Velocity component along the elongation axis, $V_x$, at $y=0$~mm. (c) Velocity component along the compression axis, $V_y$, at $x=0$~mm. (d) Magnitude of the extension rate, $|\dot\varepsilon|$, as obtained along the compression and elongation axes. The solid line is the expected relationship for a Newtonian fluid.\citep{Haward2012a} }
  \label{fgr:Fig3}
\end{figure}

The relation between $|\dot\gamma|_{(\pm0.1)}$ and the average velocity $|U|$ can therefore be established as shown in Figure~\ref{fgr:Fig2}d, leading to $|\dot\gamma|_{(\pm0.1)}=176|U|$, which satisfies the expectation for a Newtonian fluid (solid line).

The OSCER is used to generate a 2D flow field with an extensional-dominated flow in a large region around the stagnation point (the point of zero velocity at $x=y=0$, Figure~\ref{fgr:Channels}b).\citep{Haward2012a,Haward2016,Hawarda2016} The two incoming and outgoing flows are orthogonal to each other and indicated by the dashed arrows in Figure~\ref{fgr:Fig3}a. 
The $\mu$-PIV profiles for two characteristic average velocities, $|U|$, display a symmetric and Newtonian-like behavior (Figure~\ref{fgr:Fig3}a). The $x$-component of velocity, $V_x$, obtained along the elongation axis (at $y=0$) increases linearly with $x$ (Figure~\ref{fgr:Fig3}b) and, analogously, the $y$-component of velocity, $V_y$ (at $x=0$) along the compression axis, decreases linearly with $y$ (Figure~\ref{fgr:Fig3}c). 
This indicates uniform extensional rates along both the compression and elongation axes. As such, the values of compression and elongation rates can be directly obtained from the slope of the plots in Figure~\ref{fgr:Fig3}b,c. In Figure~\ref{fgr:Fig3}d, the magnitude of the compression and elongation rate as a function of the average flow velocity $|U|$ collapsed on a single curve, indicating that the extensional rates, $\dot\varepsilon$, along the compression and extension axes are equal and opposite, as for an ideal planar extension.\citep{Haward2012a} The $|\dot\varepsilon|$ followed a linear relationship with $|U|$ as $|\dot\varepsilon|=1070|U|$ (indicated by the solid line), in good agreement with previously reported Newtonian fluids.\citep{Haward2012a,Haward2016} This behavior suggests that the diluted CNC dispersion behaves as a Newtonian fluid also in extension, giving an extensional viscosity $\eta_{E}=4\eta$ where $\eta$ is the shear viscosity obtained from rheometry measurements displayed in Figure~\ref{fgr:Fig1}c and the proportionality factor of 4 is the Trouton ratio for Newtonian fluids in planar elongation flow.\citep{Hawarda2016} 

\subsection{Flow-induced alignment}
\begin{figure}[h!]
\includegraphics[width=16cm]{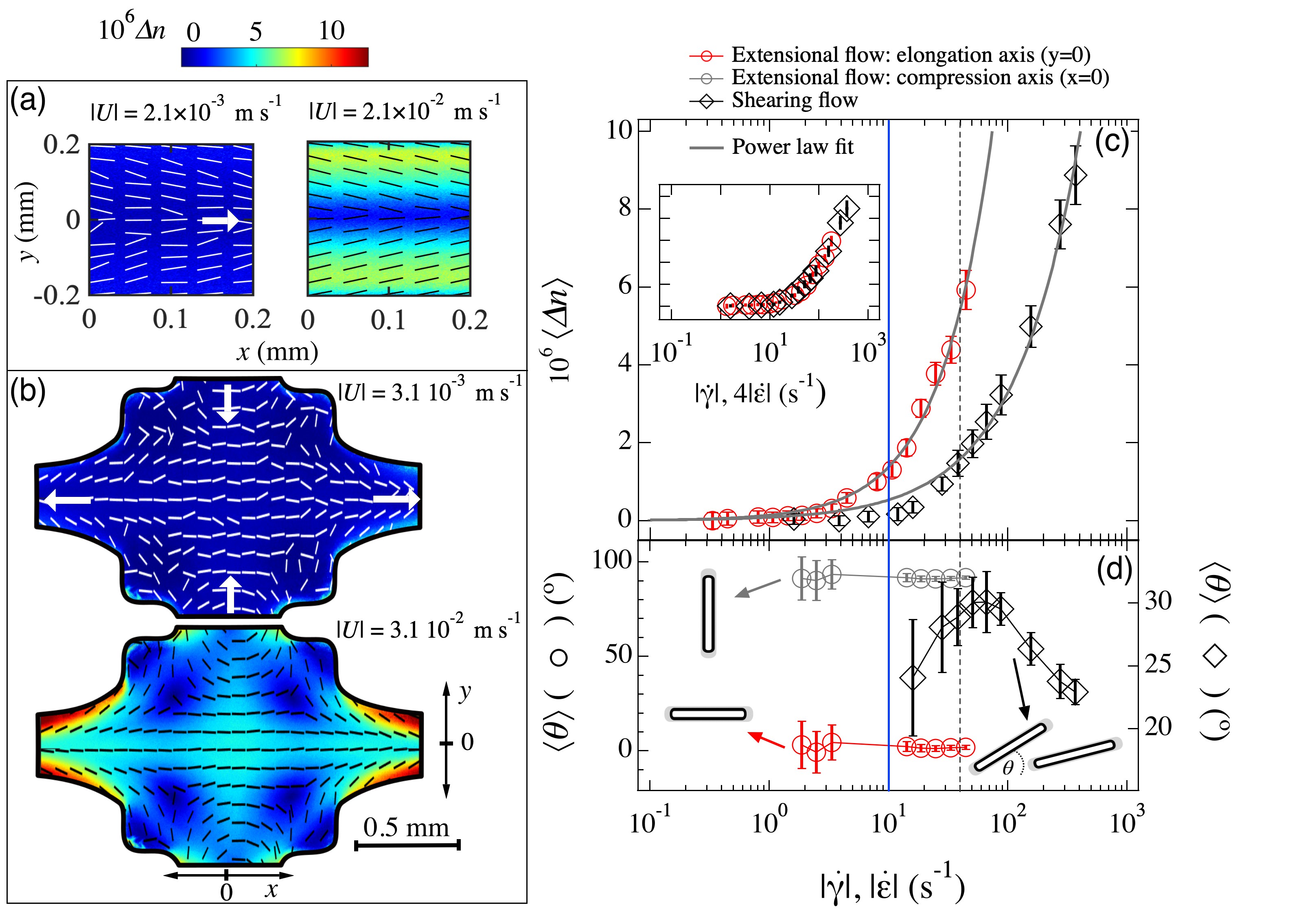}
\caption{Time averaged FIB profiles of a 0.1~wt\% CNC dispersion, for the (a) SFC and (b) OSCER device. The birefringence, $\langle \Delta n \rangle$, displayed by the contourplot whilst the direction of the slow optical axis, $\theta$, indicated by the the solid segments (in white or black, for easy visualization). (c) Spatially averaged birefringence, $\langle \Delta n \rangle$, and (d) spatially averaged orientation angle, $\langle \theta \rangle$, as a function of the magnitude of shear rate, $|\dot\gamma|$, and extension rate $|\dot\varepsilon|$. The $\langle \Delta n \rangle$ and $\langle \theta \rangle$ are obtained in the same channel location as for the relevant deformation rate. The inset in (c) displays the plot as in (c) with a re-scaled $x$-axis as $4|\dot\varepsilon|$. The solid lines in (c) are power law fittings. The dashed (black) and solid (blue) vertical lines are the values of $Dr$ and $Dr_{E}$ (s$^{-1}$) estimated \textit{via} eqn.\ref{eqn:Dr}, using values of the solvent shear viscosity ($\eta_{s}$), and solvent extensional viscosity ($\eta_{s,E}$), respectively, and $l=l_{max}$.}
  \label{fgr:Fig4}
\end{figure}

Since the spatially resolved deformation rates, $|\dot\gamma|$ and $|\dot\varepsilon|$, have been established, we shift our attention to the impact that the two different deformation rates have on the structural orientation of the dilute CNC dispersion. The contourplot in Figure~\ref{fgr:Fig4}a displays the FIB fields in the SFC at two different values of $|U|$. The birefringence intensity, $\Delta n$, is displayed by the contourplot and describes the extent of anisotropy in the system and the orientation of the slow optical axis, $\theta$, is displayed by the superimposed solid segments, which directly probes the CNC orientation angle.\citep{Vermant2001,Sun2016} As a metric of comparison, it has been recently shown that both $\Delta n$ and $\theta$ yield excellent agreement with parameters describing the extent of alignment and favourable orientation extrapolated from small angle X-ray scattering (SAXS) of anisotropic particles, enabling a trustworthy comparison of the FIB with the rheo-small angle scattering (SAS) literature.\citep{Rosen2020} The FIB fields in the SFC display the minimum at $y=0$ (where $\dot\gamma=0$~s$^{-1}$) and increases towards the channel walls (Figure~\ref{fgr:Fig4}a). Note that we attribute the narrow region of low FIB signal very close to the walls to ``shadowing'' by the wall, i.e. due to a slight imprecision of the orthogonal alignment of the microfluidic device on the imaging system (see Figure~S1 for $\Delta n$ profiles across the channel width).
Since $\Delta n$ scales with the volume fraction of aligned particles ($\phi_{aligned}$), it is clear that a greater number of CNC become aligned as the shear rate is increased. The CNC orientation, displayed by the solid segments in Figure~\ref{fgr:Fig4}a, shows an overall alignment of the particles in the flow direction. However, for the lowest value of $|U|$, the CNC orientation is not perfectly mirrored between the portion of the channel at $y<0$ mm and $y>0$ mm, and a degree of heterogeneity in the CNC orientation angle is observed. 

In the OSCER device, a large region around the stagnation point displays a strong $\Delta n$ signal, as clearly visible for the larger value of $|U|$ in Figure~\ref{fgr:Fig4}b (displayed by the light blue color in the contourplot). The direction of the CNC alignment in the compression axis is perpendicular to the flow direction ($y$-axis) due to the deceleration of the fluid element upon approaching the stagnation point and the consequent negative extensional rate along the $y$-axis. Contrarily, the CNC aligns parallel to the flow ($x$-axis) along the elongation axis due to the positive extension rate. Similar orientation trends have been reported for suspensions of anisotropic particles in extensional-dominated flows (in cross slots and fluidic four-roll mill devices), although in conditions where interparticle interactions play a crucial role on particle alignment.\citep{Pignon2003,Qazi2011}

To have a good quantitative description of the deformation rate--alignment relationship, we plot the spatially resolved values of $\Delta n$ as a function of the relevant deformation rate previously determined in the specific locations of the SFC and OSCER geometries (Figure~\ref{fgr:Fig4}c). For the SFC, the spatially averaged birefringence signal, $\langle \Delta n \rangle$, is obtained at $y=\pm0.1$ mm averaging $\Delta n$ along 1 mm in the $x$-direction, whilst for the OSCER device, $\langle \Delta n \rangle$ is obtained by averaging $\Delta n$ along 1 mm of the elongation axis (at $y=0$ over $-0.5\leq x\leq0.5$~mm, see Figure~S1, S2 for the spatially-resolved $\Delta n$ profiles). The curves showed in Figure~\ref{fgr:Fig4}c for the SFC and the OSCER device display a similar increase in $\langle \Delta n \rangle$ with the deformation rate, $|E|$, which is well captured by a power law trend as $\langle \Delta n \rangle=A |E|^{p}$, where $A$ is a proportionality factor and $p$ is the power law exponent. For the SFC, $A~=~(0.9~\pm~0.2~)~\times~10^{-7}$~s and $p=0.90 \pm 0.03$, whilst for OSCER, $A=( 1.4 \pm 0.2) \times 10^{-7} $~s  and $p=0.98 \pm 0.04$. Importantly, the FIB technique enabled to capture the CNC alignment with the deformation rate, otherwise unexpected by interpretation of the rheological measurements alone (see Figure~\ref{fgr:Fig1}c). It is noted that for a given applied $|U|$ (hence $\dot\gamma$ profile, Figure~\ref{fgr:Fig2}c), this empirical power law is able to capture the $\langle \Delta n \rangle$ profile across most of the SFC width with reasonable accuracy (see Figure~S1 for spatially-resolved $\Delta n$ profiles with the predictions of the empirical power law fitting). Interestingly, from Figure~\ref{fgr:Fig4}c, it is apparent that lower extensional rates are required to induce rod alignment in the OSCER device, than the shear rates required to induce alignment in the SFC. This appears to indicate that extensional forces are more effective at inducing CNC alignment. The $\langle \Delta n \rangle$ profiles obtained for the extensional and shearing flow collapsed on a single master curve when scaling the extensional rate as $4|\dot\varepsilon|$ (inset in Figure~\ref{fgr:Fig4}c). This indicates that extensional and shear forces have a similar relationship with $\phi_{aligned}$ and that the extensional forces are 4 times more effective for the alignment of anisotropic particles when compared to shear forces. Therefore for a Newtonian fluid containing non-interacting rods, the following proportionality can be proposed $\Delta n\propto |\dot\gamma|^{0.90} \propto 4|\dot\varepsilon|^{0.98}\propto \phi_{aligned}$. It is noted that the proportionality factor of 4 corresponds exactly to the Trouton ratio for Newtonian fluids in a planar elongation flow, which set the relationship between the solvent shear viscosity, $\eta_{s}$, and the solvent extensional viscosity, $\eta_{s,E}=4\eta_{s}$.\citep{Hawarda2016} The strong dependence of the solvent viscosity on the onset of flow alignment is reflected by the $Dr\propto 1/\eta_{s}$ in eqn.\ref{eqn:Dr}. Assuming a dominant extensional viscosity $\eta_{s,E}$ along the extensional axis, it is plausible to substitute $\eta_{s}$ with $\eta_{s,E}=4\eta_{s}$ in eqn.~\ref{eqn:Dr}, yielding a $Dr$ in extension ($Dr_{E}$) 4 times smaller compared to the $Dr$. Although not validated, the consideration of $\eta_{s}$ or $\eta_{s,E}$ for a rotational diffusion coefficient that accounts for shearing and extensional-dominated flows, respectively, has been previously proposed by Qazi et al.\citep{Qazi2011} A decreased rotational diffusion coefficient in extensional-dominated flows has also been demonstrated by Ros\'{e}n et al.\citep{Rosen2020a} for interacting cellulose nanofibrils. However, the absence of interparticle interactions and the uniform, shear-less, elongational flow presented in this work enables the first quantitative elucidation of differences between the rotational dynamics of colloidal rods in shear and extensional flows.

Considering that the longest CNC population ($l_{max}$) aligns at lower deformation rates, the predicted onset of alignment based on $Dr$ (dashed, black, lines in Figure~\ref{fgr:Fig4}c,d) and $Dr_{E}$ (solid, blue, lines in Figure~\ref{fgr:Fig4}c,d) do not exactly match the onset of birefringence. However, the scaling factor between the curves in Figure~\ref{fgr:Fig4}c strongly suggests that the larger value of $\eta_{s,E}$ compared to $\eta_{s}$ is the cause of the earlier onset of alignment.

To better evaluate the onset of alignment in shear and extensional flow, we further evaluate quantitatively the CNC orientation as captured by the angle of orientation, $\theta$, with respect to the flow direction (Figure~\ref{fgr:Fig4}d). For the SFC, the spatially averaged angle of the slow optical axis, $\langle \theta \rangle$, was determined in the same location as for the $\langle \Delta n \rangle$, i.e. averaging $|\theta|$ along 1 mm in the $x$-direction at $y=\pm 0.1$~mm, whilst for the OSCER device, the $\langle \theta \rangle$ was obtained from averaging $\theta$ along 1 mm of the elongation axis at $y=0$ (averaging $\theta$ over $-0.5\leq x\leq0.5$~mm) and compression axis at $x=0$ (averaging $\theta$ over $-0.5\leq y\leq0.5$~mm, see Figure~S2 for $\theta$ profiles along the extensional axes). In shear, for $|\dot\gamma| < 10^2$ s$^{-1}$, within experimental error (note error bars), the CNC alignment is almost constant at $\langle \theta \rangle \simeq 28 ^\circ$ with respect to the flow direction. For $|\dot\gamma|\gtrsim 10^2$ ~s$^{-1}$, $\langle \theta \rangle$ progressively decreases to values of around $22 ^\circ$ at the highest shear rates, in good agreement with SAXS and birefringence studies of a diluted CNC suspension.\citep{Rosen2020}
Along the elongation axis of the OSCER, the CNC aligns parallel to the flow, with an orientation of $\langle \theta \rangle \simeq 0 ^\circ$ to the flow direction. Contrarily, along the compression axis, the CNC align perpendicular to the flow with $\langle \theta \rangle \simeq 90 ^\circ$. Such perpendicular alignment of elongated particles has been previously reported for shear thinning fluids in extensional-dominated flows.\citep{Kiriya2012,Trebbin2013} However, it is clear from our results that shear thinning is not a requirement for such orientation to occur.

As previously anticipated, the expected value for the onset of CNC alignment is $|\dot\gamma|\simeq~40$~s$^{-1}$ (depicted by $Dr$, dashed (black) lines in Figure~\ref{fgr:Fig4}c,d) when considering the longest CNC population. Although the onset of birefringence, $\langle \Delta n \rangle$, occurs at values of $|\dot\gamma|\approx 10$~s$^{-1}$, the predicted onset of alignment at $|\dot\gamma|\simeq 40$~s$^{-1}$ is in close agreement with the origin of the significant decrease in $\langle \theta \rangle$ in the SFC (Figure~\ref{fgr:Fig4}d). The coexistence of $\langle \Delta n \rangle > 0$ and the steady value of $\langle \theta \rangle$ with a relatively large error at $|\dot\gamma| \lesssim 40$ s$^{-1}$, suggests the presence of an intermediate shear rate region where Brownian diffusion and deformation rate are of comparable magnitude, leading to a time-dependent and collective particle rotation in the velocity-gradient plane.\citep{Dhont2007,Mewis1997} Contrarily, a Brownian dominated region occurs for $|\dot\gamma| < 10$ s$^{-1}$, where the system is isotropic and $\langle \Delta n \rangle\approx 0$, whilst, at $|\dot\gamma| \gtrsim 40$ s$^{-1}$, the hydrodynamic forces orient the particles towards a $\langle \theta \rangle = 0^\circ$, causing a pronounced increase in $\langle \Delta n \rangle$. A similar intermediate shear rate regime has been described for a concentrated CNC suspension\citep{Ebeling1999} as well as wormlike micelles\citep{Mu2019} and polymeric liquid crystals.\citep{Mewis1997}

For the elongational flow, $\langle \theta \rangle$ is constant over the whole range of $|\dot\varepsilon|$, in agreement with the reported analytical solutions for diluted Brownian suspensions of rod-like particles.\citep{Dhont2007} 

Therefore, the onset of a convection-dominated particle alignment in a shear-dominated flow is more adequately described by the orientation angle $\langle \theta \rangle$, whilst, for the extensional-dominated flow, the onset of birefringence is a better parameter to consider since $\langle \theta \rangle$ is independent of the rate of deformation. When this is considered, the onset of alignment is satisfactorily captured by a diffusion coefficient that accounts for the longest particle population and the appropriate viscosity that the fluid experiences in the specific location of the channel, i.e. $Dr$ or $Dr_{E}$.

\section{CONCLUSION}
Combining flow visualization and flow induced birefringence, we are able to decouple the effects of shear and extensional forces on a dilute dispersion of cellulose nanocrystals, consisting of negatively charged polydisperse and rod-like rigid particles. 
Simultaneous analysis of the birefringence with the orientation of the slow optical axis elucidate that extensional forces are \textit{ca.} 4 times more effective at inducing particle alignment compared to shear forces. This difference is explained by the different viscosity that the fluid experiences between a shear and extensional-dominated region, namely the shear and extensional viscosity, respectively. In addition, shear and extension rates have displayed a remarkably different effect on particle orientation. In a shear-dominated flow, the particles exhibit a gradual decrease in the orientation angle with increasing shear rate. Contrarily, in the extensional-dominated flow the particle orientation is deformation-rate independent and a parallel or perpendicular orientation of the particles to the flow direction is observed along the elongation and compression axes of the geometry, respectively. To the best of our knowledge, this is the first experimental report where the effects of shear and pure extensional flows on the alignment of rigid anisotropic particles are quantitatively compared. 
This understanding will provide the basis to decouple the orientation dynamics in industrially relevant flows, where a combination of shear and extensional rates are usually present and in fluids where interparticle interactions play a crucial role. We envisage that the knowledge provided in this work will help to optimize engineering processes involved with controlled anisotropy. Having demonstrated the influence of the solvent shear and extensional viscosity at controlling the alignment of non-interacting rods for a Newtonian solvent, we aim in future work to broaden this understanding for the more complex non-Newtonian case, by using solvent fluids with larger and deformation rate--dependent Trouton ratios. 

\begin{acknowledgement}
The authors gratefully acknowledge the support of Okinawa Institute of Science and Technology Graduate University with subsidy funding from the Cabinet Office, Government of Japan. S.J.H. and A.Q.S. also acknowledge financial support from the Japanese Society for the Promotion of Science (JSPS, Grant Nos. 18K03958 and 18H01135) and the Joint Research Projects (JRPs) supported by the JSPS and the Swiss National Science Foundation (SNSF). The authors thank Dr.~Riccardo Funari from the Micro/Bio/Nanofluidics unit at OIST for his assistance with AFM measurements and Dr.~Vikram Rathee from the Micro/Bio/Nanofluidics unit at OIST for comments on the manuscript.
\end{acknowledgement}


\providecommand{\latin}[1]{#1}
\makeatletter
\providecommand{\doi}
  {\begingroup\let\do\@makeother\dospecials
  \catcode`\{=1 \catcode`\}=2 \doi@aux}
\providecommand{\doi@aux}[1]{\endgroup\texttt{#1}}
\makeatother
\providecommand*\mcitethebibliography{\thebibliography}
\csname @ifundefined\endcsname{endmcitethebibliography}
  {\let\endmcitethebibliography\endthebibliography}{}


\begin{mcitethebibliography}{45}
\providecommand*\natexlab[1]{#1}
\providecommand*\mciteSetBstSublistMode[1]{}
\providecommand*\mciteSetBstMaxWidthForm[2]{}
\providecommand*\mciteBstWouldAddEndPuncttrue
  {\def\EndOfBibitem{\unskip.}}
\providecommand*\mciteBstWouldAddEndPunctfalse
  {\let\EndOfBibitem\relax}
\providecommand*\mciteSetBstMidEndSepPunct[3]{}
\providecommand*\mciteSetBstSublistLabelBeginEnd[3]{}
\providecommand*\EndOfBibitem{}
\mciteSetBstSublistMode{f}
\mciteSetBstMaxWidthForm{subitem}{(\alph{mcitesubitemcount})}
\mciteSetBstSublistLabelBeginEnd
  {\mcitemaxwidthsubitemform\space}
  {\relax}
  {\relax}

\bibitem[Solomon and Spicer(2010)Solomon, and Spicer]{Solomon2010}
Solomon,~M.~J.; Spicer,~P.~T. {Microstructural regimes of colloidal rod
  suspensions, gels, and glasses}. \emph{Soft Matter} \textbf{2010}, \emph{6},
  1391--1400\relax
\mciteBstWouldAddEndPuncttrue
\mciteSetBstMidEndSepPunct{\mcitedefaultmidpunct}
{\mcitedefaultendpunct}{\mcitedefaultseppunct}\relax
\EndOfBibitem
\bibitem[Ros{\'{e}}n \latin{et~al.}(2020)Ros{\'{e}}n, Hsiao, and
  S{\"{o}}derberg]{Rosen2020b}
Ros{\'{e}}n,~T.; Hsiao,~B.~S.; S{\"{o}}derberg,~L.~D. {Elucidating the
  opportunities and challenges for nanocellulose spinning}. \emph{Advanced
  Materials} \textbf{2020}, 2001238\relax
\mciteBstWouldAddEndPuncttrue
\mciteSetBstMidEndSepPunct{\mcitedefaultmidpunct}
{\mcitedefaultendpunct}{\mcitedefaultseppunct}\relax
\EndOfBibitem
\bibitem[Calabrese \latin{et~al.}(2020)Calabrese, da~Silva, Porcar, Bryant,
  Hossain, Scott, and Edler]{Calabrese2020b}
Calabrese,~V.; da~Silva,~M.~A.; Porcar,~L.; Bryant,~S.~J.; Hossain,~K. M.~Z.;
  Scott,~J.~L.; Edler,~K.~J. {Filler size effect in an attractive fibrillated
  network: a structural and rheological perspective}. \emph{Soft Matter}
  \textbf{2020}, \emph{16}, 3303--3310\relax
\mciteBstWouldAddEndPuncttrue
\mciteSetBstMidEndSepPunct{\mcitedefaultmidpunct}
{\mcitedefaultendpunct}{\mcitedefaultseppunct}\relax
\EndOfBibitem
\bibitem[H{\aa}kansson \latin{et~al.}(2014)H{\aa}kansson, Fall, Lundell, Yu,
  Krywka, Roth, Santoro, Kvick, {Prahl Wittberg}, W{\aa}gberg, and
  S{\"{o}}derberg]{Hakansson2014}
H{\aa}kansson,~K. M.~O.; Fall,~A.~B.; Lundell,~F.; Yu,~S.; Krywka,~C.;
  Roth,~S.~V.; Santoro,~G.; Kvick,~M.; {Prahl Wittberg},~L.; W{\aa}gberg,~L.;
  S{\"{o}}derberg,~L.~D. {Hydrodynamic alignment and assembly of nanofibrils
  resulting in strong cellulose filaments}. \emph{Nature Communications}
  \textbf{2014}, \emph{5}, 4018\relax
\mciteBstWouldAddEndPuncttrue
\mciteSetBstMidEndSepPunct{\mcitedefaultmidpunct}
{\mcitedefaultendpunct}{\mcitedefaultseppunct}\relax
\EndOfBibitem
\bibitem[Nechyporchuk \latin{et~al.}(2018)Nechyporchuk, H{\aa}kansson, Gowda.V,
  Lundell, Hagstr{\"{o}}m, and K{\"{o}}hnke]{Nechyporchuk2019}
Nechyporchuk,~O.; H{\aa}kansson,~K. M.~O.; Gowda.V,~K.; Lundell,~F.;
  Hagstr{\"{o}}m,~B.; K{\"{o}}hnke,~T. {Continuous assembly of cellulose
  nanofibrils and nanocrystals into strong macrofibers through microfluidic
  spinning}. \emph{Advanced Materials Technologies} \textbf{2018}, \emph{4},
  1800557\relax
\mciteBstWouldAddEndPuncttrue
\mciteSetBstMidEndSepPunct{\mcitedefaultmidpunct}
{\mcitedefaultendpunct}{\mcitedefaultseppunct}\relax
\EndOfBibitem
\bibitem[Liu \latin{et~al.}(2014)Liu, Wang, Ma, Tian, Gu, and Lin]{Liu2014}
Liu,~D.; Wang,~S.; Ma,~Z.; Tian,~D.; Gu,~M.; Lin,~F. {Structure-color mechanism
  of iridescent cellulose nanocrystal films}. \emph{RSC Advances}
  \textbf{2014}, \emph{4}, 39322--39331\relax
\mciteBstWouldAddEndPuncttrue
\mciteSetBstMidEndSepPunct{\mcitedefaultmidpunct}
{\mcitedefaultendpunct}{\mcitedefaultseppunct}\relax
\EndOfBibitem
\bibitem[Kiriya \latin{et~al.}(2012)Kiriya, Kawano, Onoe, and
  Takeuchi]{Kiriya2012}
Kiriya,~D.; Kawano,~R.; Onoe,~H.; Takeuchi,~S. {Microfluidic control of the
  internal morphology in nanofiber-based macroscopic cables}. \emph{Angewandte
  Chemie - International Edition} \textbf{2012}, \emph{51}, 7942--7947\relax
\mciteBstWouldAddEndPuncttrue
\mciteSetBstMidEndSepPunct{\mcitedefaultmidpunct}
{\mcitedefaultendpunct}{\mcitedefaultseppunct}\relax
\EndOfBibitem
\bibitem[Xin \latin{et~al.}(2019)Xin, Zhu, Deng, Cheng, Zhang, Chung, De, and
  Lian]{Xin2019}
Xin,~G.; Zhu,~W.; Deng,~Y.; Cheng,~J.; Zhang,~L.~T.; Chung,~A.~J.; De,~S.;
  Lian,~J. {Microfluidics-enabled orientation and microstructure control of
  macroscopic graphene fibres}. \emph{Nature Nanotechnology} \textbf{2019},
  \emph{14}, 168--175\relax
\mciteBstWouldAddEndPuncttrue
\mciteSetBstMidEndSepPunct{\mcitedefaultmidpunct}
{\mcitedefaultendpunct}{\mcitedefaultseppunct}\relax
\EndOfBibitem
\bibitem[{De France} \latin{et~al.}(2017){De France}, Yager, Chan, Corbett,
  Cranston, and Hoare]{DeFrance2017}
{De France},~K.~J.; Yager,~K.~G.; Chan,~K. J.~W.; Corbett,~B.; Cranston,~E.~D.;
  Hoare,~T. {Injectable anisotropic nanocomposite hydrogels direct in situ
  growth and alignment of myotubes}. \emph{Nano Letters} \textbf{2017},
  \emph{17}, 6487--6495\relax
\mciteBstWouldAddEndPuncttrue
\mciteSetBstMidEndSepPunct{\mcitedefaultmidpunct}
{\mcitedefaultendpunct}{\mcitedefaultseppunct}\relax
\EndOfBibitem
\bibitem[Doi and Edwards(1978)Doi, and Edwards]{LangStiffness}
Doi,~M.; Edwards,~S.~F. {Dynamics of rod-like macromolecules in concentrated
  solution. Part 1—{B}rownian motion in the equilibrium state }. \emph{Journal of the Chemical Society, Faraday
  Transactions 2: Molecular and Chemical Physics} \textbf{1978}, \emph{74},
  560--570\relax
\mciteBstWouldAddEndPuncttrue
\mciteSetBstMidEndSepPunct{\mcitedefaultmidpunct}
{\mcitedefaultendpunct}{\mcitedefaultseppunct}\relax
\EndOfBibitem
\bibitem[Vermant \latin{et~al.}(2001)Vermant, Yang, and Fuller]{Vermant2001}
Vermant,~J.; Yang,~H.; Fuller,~G.~G. {Rheo-optical determination of aspect ratio
  and polydispersity of nonspherical particles}. \emph{AIChE Journal}
  \textbf{2001}, \emph{47}, 790--798\relax
\mciteBstWouldAddEndPuncttrue
\mciteSetBstMidEndSepPunct{\mcitedefaultmidpunct}
{\mcitedefaultendpunct}{\mcitedefaultseppunct}\relax
\EndOfBibitem
\bibitem[Dhont and Briels(2007)Dhont, and Briels]{Dhont2007}
Dhont,~J. K.~G.; Briels,~W.~J. \emph{Soft Matter Vol. 2}; Wiley-VCH: Weinheim,
  Germany, 2007; Vol.~2; pp 216--283\relax
\mciteBstWouldAddEndPuncttrue
\mciteSetBstMidEndSepPunct{\mcitedefaultmidpunct}
{\mcitedefaultendpunct}{\mcitedefaultseppunct}\relax
\EndOfBibitem
\bibitem[Reddy \latin{et~al.}(2018)Reddy, Natale, Prud'homme, and
  Vermant]{Reddy2018}
Reddy,~N.~K.; Natale,~G.; Prud'homme,~R.~K.; Vermant,~J. {Rheo-optical analysis
  of functionalized graphene suspensions}. \emph{Langmuir} \textbf{2018},
  \emph{34}, 7844--7851\relax
\mciteBstWouldAddEndPuncttrue
\mciteSetBstMidEndSepPunct{\mcitedefaultmidpunct}
{\mcitedefaultendpunct}{\mcitedefaultseppunct}\relax
\EndOfBibitem
\bibitem[Winkler \latin{et~al.}(2004)Winkler, Mussawisade, Ripoll, and
  Gompper]{Winkler2004}
Winkler,~R.~G.; Mussawisade,~K.; Ripoll,~M.; Gompper,~G. {Rod-like colloids and
  polymers in shear flow: a multi-particle-collision dynamics study}.
  \emph{Journal of Physics: Condensed Matter} \textbf{2004}, \emph{16},
  3941--3954\relax
\mciteBstWouldAddEndPuncttrue
\mciteSetBstMidEndSepPunct{\mcitedefaultmidpunct}
{\mcitedefaultendpunct}{\mcitedefaultseppunct}\relax
\EndOfBibitem
\bibitem[Haward \latin{et~al.}(2016)Haward, McKinley, and Shen]{Haward2016}
Haward,~S.~J.; McKinley,~G.~H.; Shen,~A.~Q. {Elastic instabilities in planar
  elongational flow of monodisperse polymer solutions}. \emph{Scientific
  Reports} \textbf{2016}, \emph{6}, 33029\relax
\mciteBstWouldAddEndPuncttrue
\mciteSetBstMidEndSepPunct{\mcitedefaultmidpunct}
{\mcitedefaultendpunct}{\mcitedefaultseppunct}\relax
\EndOfBibitem
\bibitem[Lang \latin{et~al.}(2019)Lang, Hendricks, Zhang, Reddy, Rothstein,
  Lettinga, Vermant, and Clasen]{Lang2019a}
Lang,~C.; Hendricks,~J.; Zhang,~Z.; Reddy,~N.~K.; Rothstein,~J.~P.;
  Lettinga,~M.~P.; Vermant,~J.; Clasen,~C. {Effects of particle stiffness on
  the extensional rheology of model rod-like nanoparticle suspensions}.
  \emph{Soft Matter} \textbf{2019}, \emph{15}, 833--841\relax
\mciteBstWouldAddEndPuncttrue
\mciteSetBstMidEndSepPunct{\mcitedefaultmidpunct}
{\mcitedefaultendpunct}{\mcitedefaultseppunct}\relax
\EndOfBibitem
\bibitem[Trebbin \latin{et~al.}(2013)Trebbin, Steinhauser, Perlich, Buffet,
  Roth, Zimmermann, Thiele, and F{\"{o}}rster]{Trebbin2013}
Trebbin,~M.; Steinhauser,~D.; Perlich,~J.; Buffet,~A.; Roth,~S.~V.;
  Zimmermann,~W.; Thiele,~J.; F{\"{o}}rster,~S. {Anisotropic particles align
  perpendicular to the flow direction in narrow microchannels}.
  \emph{Proceedings of the National Academy of Sciences of the United States of
  America} \textbf{2013}, \emph{110}, 6706--6711\relax
\mciteBstWouldAddEndPuncttrue
\mciteSetBstMidEndSepPunct{\mcitedefaultmidpunct}
{\mcitedefaultendpunct}{\mcitedefaultseppunct}\relax
\EndOfBibitem
\bibitem[Qazi \latin{et~al.}(2011)Qazi, Rennie, Tucker, Penfold, and
  Grillo]{Qazi2011}
Qazi,~S. J.~S.; Rennie,~A.~R.; Tucker,~I.; Penfold,~J.; Grillo,~I. {Alignment
  of dispersions of plate-like colloidal particles of Ni(OH)$_2$ induced by
  elongational flow}. \emph{Journal of Physical Chemistry B} \textbf{2011},
  \emph{115}, 3271--3280\relax
\mciteBstWouldAddEndPuncttrue
\mciteSetBstMidEndSepPunct{\mcitedefaultmidpunct}
{\mcitedefaultendpunct}{\mcitedefaultseppunct}\relax
\EndOfBibitem
\bibitem[Pignon \latin{et~al.}(2003)Pignon, Magnin, Piau, and
  Fuller]{Pignon2003}
Pignon,~F.; Magnin,~A.; Piau,~J.-M.; Fuller,~G.~G. {The orientation dynamics of
  rigid rod suspensions under extensional flow}. \emph{Journal of Rheology}
  \textbf{2003}, \emph{47}, 371--388\relax
\mciteBstWouldAddEndPuncttrue
\mciteSetBstMidEndSepPunct{\mcitedefaultmidpunct}
{\mcitedefaultendpunct}{\mcitedefaultseppunct}\relax
\EndOfBibitem
\bibitem[Corona \latin{et~al.}(2018)Corona, Ruocco, Weigandt, Leal, and
  Helgeson]{Corona2018}
Corona,~P.~T.; Ruocco,~N.; Weigandt,~K.~M.; Leal,~L.~G.; Helgeson,~M.~E.
  {Probing flow-induced nanostructure of complex fluids in arbitrary 2D flows
  using a fluidic four-roll mill (FFoRM)}. \emph{Scientific Reports}
  \textbf{2018}, \emph{8}, 15559\relax
\mciteBstWouldAddEndPuncttrue
\mciteSetBstMidEndSepPunct{\mcitedefaultmidpunct}
{\mcitedefaultendpunct}{\mcitedefaultseppunct}\relax
\EndOfBibitem
\bibitem[Ros{\'{e}}n \latin{et~al.}(2020)Ros{\'{e}}n, Mittal, Roth, Zhang,
  Lundell, and S{\"{o}}derberg]{Rosen2020a}
Ros{\'{e}}n,~T.; Mittal,~N.; Roth,~S.~V.; Zhang,~P.; Lundell,~F.;
  S{\"{o}}derberg,~L.~D. {Flow fields control nanostructural organization in
  semiflexible networks}. \emph{Soft Matter} \textbf{2020}, \emph{16},
  5439--5449\relax
\mciteBstWouldAddEndPuncttrue
\mciteSetBstMidEndSepPunct{\mcitedefaultmidpunct}
{\mcitedefaultendpunct}{\mcitedefaultseppunct}\relax
\EndOfBibitem
\bibitem[Hasegawa \latin{et~al.}(2020)Hasegawa, Horikawa, and
  Shikata]{Hasegawa2020a}
Hasegawa,~H.; Horikawa,~Y.; Shikata,~T. {Cellulose nanocrystals as a model
  substance for rigid rod particle suspension rheology}. \emph{Macromolecules}
  \textbf{2020}, \emph{53}, 2677--2685\relax
\mciteBstWouldAddEndPuncttrue
\mciteSetBstMidEndSepPunct{\mcitedefaultmidpunct}
{\mcitedefaultendpunct}{\mcitedefaultseppunct}\relax
\EndOfBibitem
\bibitem[Bertsch \latin{et~al.}(2017)Bertsch, Isabettini, and
  Fischer]{Bertsch2017}
Bertsch,~P.; Isabettini,~S.; Fischer,~P. {Ion-induced hydrogel formation and
  nematic ordering of nanocrystalline cellulose suspensions}.
  \emph{Biomacromolecules} \textbf{2017}, \emph{18}, 4060--4066\relax
\mciteBstWouldAddEndPuncttrue
\mciteSetBstMidEndSepPunct{\mcitedefaultmidpunct}
{\mcitedefaultendpunct}{\mcitedefaultseppunct}\relax
\EndOfBibitem
\bibitem[Bertsch \latin{et~al.}(2019)Bertsch, S{\'{a}}nchez-Ferrer, Bagnani,
  Isabettini, Kohlbrecher, Mezzenga, and Fischer]{Bertsch2019}
Bertsch,~P.; S{\'{a}}nchez-Ferrer,~A.; Bagnani,~M.; Isabettini,~S.;
  Kohlbrecher,~J.; Mezzenga,~R.; Fischer,~P. {Ion-induced formation of
  nanocrystalline cellulose colloidal glasses containing nematic domains}.
  \emph{Langmuir} \textbf{2019}, \emph{35}, 4117--4124\relax
\mciteBstWouldAddEndPuncttrue
\mciteSetBstMidEndSepPunct{\mcitedefaultmidpunct}
{\mcitedefaultendpunct}{\mcitedefaultseppunct}\relax
\EndOfBibitem
\bibitem[Reid \latin{et~al.}(2017)Reid, Villalobos, and Cranston]{Reid2017}
Reid,~M.~S.; Villalobos,~M.; Cranston,~E.~D. {Benchmarking cellulose
  nanocrystals: From the laboratory to industrial production}. \emph{Langmuir}
  \textbf{2017}, \emph{33}, 1583--1598\relax
\mciteBstWouldAddEndPuncttrue
\mciteSetBstMidEndSepPunct{\mcitedefaultmidpunct}
{\mcitedefaultendpunct}{\mcitedefaultseppunct}\relax
\EndOfBibitem
\bibitem[Usov and Mezzenga(2015)Usov, and Mezzenga]{Usov2015a}
Usov,~I.; Mezzenga,~R. {FiberApp: An open-source software for tracking and
  analyzing polymers, filaments, biomacromolecules, and fibrous objects}.
  \emph{Macromolecules} \textbf{2015}, \emph{48}, 1269--1280\relax
\mciteBstWouldAddEndPuncttrue
\mciteSetBstMidEndSepPunct{\mcitedefaultmidpunct}
{\mcitedefaultendpunct}{\mcitedefaultseppunct}\relax
\EndOfBibitem
\bibitem[Haward \latin{et~al.}(2012)Haward, Oliveira, Alves, and
  McKinley]{Haward2012a}
Haward,~S.~J.; Oliveira,~M. S.~N.; Alves,~M.~A.; McKinley,~G.~H. {Optimized
  cross-slot flow geometry for microfluidic extensional rheometry}.
  \emph{Physical Review Letters} \textbf{2012}, \emph{109}, 128301\relax
\mciteBstWouldAddEndPuncttrue
\mciteSetBstMidEndSepPunct{\mcitedefaultmidpunct}
{\mcitedefaultendpunct}{\mcitedefaultseppunct}\relax
\EndOfBibitem
\bibitem[Burshtein \latin{et~al.}(2019)Burshtein, Chan, Toda-Peters, Shen, and
  Haward]{Burshtein2019}
Burshtein,~N.; Chan,~S.~T.; Toda-Peters,~K.; Shen,~A.~Q.; Haward,~S.~J.
  {3D-printed glass microfluidics for fluid dynamics and rheology}.
  \emph{Current Opinion in Colloid and Interface Science} \textbf{2019},
  \emph{43}, 1--14\relax
\mciteBstWouldAddEndPuncttrue
\mciteSetBstMidEndSepPunct{\mcitedefaultmidpunct}
{\mcitedefaultendpunct}{\mcitedefaultseppunct}\relax
\EndOfBibitem
\bibitem[Haward \latin{et~al.}(2019)Haward, Kitajima, Toda-Peters, Takahashi,
  and Shen]{Haward2019}
Haward,~S.~J.; Kitajima,~N.; Toda-Peters,~K.; Takahashi,~T.; Shen,~A.~Q. {Flow
  of wormlike micellar solutions around microfluidic cylinders with high aspect
  ratio and low blockage ratio}. \emph{Soft Matter} \textbf{2019}, \emph{15},
  1927--1941\relax
\mciteBstWouldAddEndPuncttrue
\mciteSetBstMidEndSepPunct{\mcitedefaultmidpunct}
{\mcitedefaultendpunct}{\mcitedefaultseppunct}\relax
\EndOfBibitem
\bibitem[Meinhart \latin{et~al.}(2000)Meinhart, Wereley, and
  Gray]{Meinhart2000}
Meinhart,~C.~D.; Wereley,~S.; Gray,~M. {Volume illumination for two-dimensional
  particle image velocimetry}. \emph{Measurement Science and Technology}
  \textbf{2000}, \emph{11}, 809--814\relax
\mciteBstWouldAddEndPuncttrue
\mciteSetBstMidEndSepPunct{\mcitedefaultmidpunct}
{\mcitedefaultendpunct}{\mcitedefaultseppunct}\relax
\EndOfBibitem
\bibitem[Wagner \latin{et~al.}(2010)Wagner, Raman, and
  Moon]{RyanWagnerArvindRaman2010}
Wagner,~R.; Raman,~A.; Moon,~R.~J. \emph{10th International Conference on Wood
  {\&} Biofiber Plastic Composites}; 2010; pp 309--316\relax
\mciteBstWouldAddEndPuncttrue
\mciteSetBstMidEndSepPunct{\mcitedefaultmidpunct}
{\mcitedefaultendpunct}{\mcitedefaultseppunct}\relax
\EndOfBibitem
\bibitem[Lang \latin{et~al.}(2019)Lang, Kohlbrecher, Porcar, Radulescu,
  Sellinghoff, Dhont, and Lettinga]{Lang2019}
Lang,~C.; Kohlbrecher,~J.; Porcar,~L.; Radulescu,~A.; Sellinghoff,~K.;
  Dhont,~J. K.~G.; Lettinga,~M.~P. {Microstructural understanding of the
  length- and stiffness-dependent shear thinning in semidilute colloidal rods}.
  \emph{Macromolecules} \textbf{2019}, \emph{52}, 9604--9612\relax
\mciteBstWouldAddEndPuncttrue
\mciteSetBstMidEndSepPunct{\mcitedefaultmidpunct}
{\mcitedefaultendpunct}{\mcitedefaultseppunct}\relax
\EndOfBibitem
\bibitem[Usov \latin{et~al.}(2015)Usov, Nystr{\"{o}}m, Adamcik, Handschin,
  Sch{\"{u}}tz, Fall, Bergstr{\"{o}}m, and Mezzenga]{Usov2015}
Usov,~I.; Nystr{\"{o}}m,~G.; Adamcik,~J.; Handschin,~S.; Sch{\"{u}}tz,~C.;
  Fall,~A.; Bergstr{\"{o}}m,~L.; Mezzenga,~R. {Understanding nanocellulose
  chirality and structure–properties relationship at the single fibril
  level}. \emph{Nature Communications} \textbf{2015}, \emph{6}, 7564\relax
\mciteBstWouldAddEndPuncttrue
\mciteSetBstMidEndSepPunct{\mcitedefaultmidpunct}
{\mcitedefaultendpunct}{\mcitedefaultseppunct}\relax
\EndOfBibitem
\bibitem[Lang and Lettinga(2020)Lang, and Lettinga]{Lang2020b}
Lang,~C.; Lettinga,~M.~P. {Shear flow behavior of bidisperse rodlike colloids}.
  \emph{Macromolecules} \textbf{2020}, \emph{53}, 2662--2668\relax
\mciteBstWouldAddEndPuncttrue
\mciteSetBstMidEndSepPunct{\mcitedefaultmidpunct}
{\mcitedefaultendpunct}{\mcitedefaultseppunct}\relax
\EndOfBibitem
\bibitem[Tanaka \latin{et~al.}(2014)Tanaka, Saito, Ishii, and
  Isogai]{Tanaka2014}
Tanaka,~R.; Saito,~T.; Ishii,~D.; Isogai,~A. {Determination of nanocellulose
  fibril length by shear viscosity measurement}. \emph{Cellulose}
  \textbf{2014}, \emph{21}, 1581--1589\relax
\mciteBstWouldAddEndPuncttrue
\mciteSetBstMidEndSepPunct{\mcitedefaultmidpunct}
{\mcitedefaultendpunct}{\mcitedefaultseppunct}\relax
\EndOfBibitem
\bibitem[Kobayashi and Yamamoto(2011)Kobayashi, and Yamamoto]{Kobayashi2011}
Kobayashi,~H.; Yamamoto,~R. {Reentrant transition in the shear viscosity of
  dilute rigid-rod dispersions}. \emph{Physical Review E} \textbf{2011},
  \emph{84}, 051404\relax
\mciteBstWouldAddEndPuncttrue
\mciteSetBstMidEndSepPunct{\mcitedefaultmidpunct}
{\mcitedefaultendpunct}{\mcitedefaultseppunct}\relax
\EndOfBibitem
\bibitem[Lang \latin{et~al.}(2016)Lang, Kohlbrecher, Porcar, and
  Lettinga]{Lang2016}
Lang,~C.; Kohlbrecher,~J.; Porcar,~L.; Lettinga,~M. {The connection between
  biaxial orientation and shear thinning for quasi-ideal rods}. \emph{Polymers}
  \textbf{2016}, \emph{8}, 291\relax
\mciteBstWouldAddEndPuncttrue
\mciteSetBstMidEndSepPunct{\mcitedefaultmidpunct}
{\mcitedefaultendpunct}{\mcitedefaultseppunct}\relax
\EndOfBibitem
\bibitem[Shah and London(1978)Shah, and London]{Shah}
Shah,~R.; London,~A. In \emph{{Laminar Flow Forced Convection in Ducts: a
  Source Book for Compact Heat Exchanger Analytical Data, Academic Press, New
  York, 1978.}}; Hartnett,~T. I. J.~P., Ed.; Academic Press, 1978\relax
\mciteBstWouldAddEndPuncttrue
\mciteSetBstMidEndSepPunct{\mcitedefaultmidpunct}
{\mcitedefaultendpunct}{\mcitedefaultseppunct}\relax
\EndOfBibitem
\bibitem[Haward(2016)]{Hawarda2016}
Haward,~S.~J. {Microfluidic extensional rheometry using stagnation point flow}.
  \emph{Biomicrofluidics} \textbf{2016}, \emph{10}, 043401\relax
\mciteBstWouldAddEndPuncttrue
\mciteSetBstMidEndSepPunct{\mcitedefaultmidpunct}
{\mcitedefaultendpunct}{\mcitedefaultseppunct}\relax
\EndOfBibitem
\bibitem[Sun and Huang(2016)Sun, and Huang]{Sun2016}
Sun,~C.-L.; Huang,~H.-Y. {Measurements of flow-induced birefringence in
  microfluidics}. \emph{Biomicrofluidics} \textbf{2016}, \emph{10},
  011903\relax
\mciteBstWouldAddEndPuncttrue
\mciteSetBstMidEndSepPunct{\mcitedefaultmidpunct}
{\mcitedefaultendpunct}{\mcitedefaultseppunct}\relax
\EndOfBibitem
\bibitem[Ros{\'{e}}n \latin{et~al.}(2020)Ros{\'{e}}n, Wang, Zhan, He,
  Chodankar, and Hsiao]{Rosen2020}
Ros{\'{e}}n,~T.; Wang,~R.; Zhan,~C.; He,~H.; Chodankar,~S.; Hsiao,~B.~S.
  {Cellulose nanofibrils and nanocrystals in confined flow: Single-particle
  dynamics to collective alignment revealed through scanning small-angle X-ray
  scattering and numerical simulations}. \emph{Physical Review E}
  \textbf{2020}, \emph{101}, 032610\relax
\mciteBstWouldAddEndPuncttrue
\mciteSetBstMidEndSepPunct{\mcitedefaultmidpunct}
{\mcitedefaultendpunct}{\mcitedefaultseppunct}\relax
\EndOfBibitem
\bibitem[Mewis \latin{et~al.}(1997)Mewis, Mortier, Vermant, and
  Moldenaers]{Mewis1997}
Mewis,~J.; Mortier,~M.; Vermant,~J.; Moldenaers,~P. {Experimental evidence for
  the existence of a wagging regime in polymeric liquid crystals}.
  \emph{Macromolecules} \textbf{1997}, \emph{30}, 1323--1328\relax
\mciteBstWouldAddEndPuncttrue
\mciteSetBstMidEndSepPunct{\mcitedefaultmidpunct}
{\mcitedefaultendpunct}{\mcitedefaultseppunct}\relax
\EndOfBibitem
\bibitem[Ebeling \latin{et~al.}(1999)Ebeling, Paillet, Borsali, Diat, Dufresne,
  Cavaill{\'{e}}, and Chanzy]{Ebeling1999}
Ebeling,~T.; Paillet,~M.; Borsali,~R.; Diat,~O.; Dufresne,~A.;
  Cavaill{\'{e}},~J.~Y.; Chanzy,~H. {Shear-induced orientation phenomena in
  suspensions of cellulose microcrystals, revealed by small angle X-ray
  scattering}. \emph{Langmuir} \textbf{1999}, \emph{15}, 6123--6126\relax
\mciteBstWouldAddEndPuncttrue
\mciteSetBstMidEndSepPunct{\mcitedefaultmidpunct}
{\mcitedefaultendpunct}{\mcitedefaultseppunct}\relax
\EndOfBibitem
\bibitem[Lerouge and Berret(2009)Lerouge, and Berret]{Mu2019}
Lerouge,~S.; Berret,~J.-F. In \emph{Polymer Characterization: Rheology, Laser
  Interferometry, Electrooptics}; Dusek,~K., Joanny,~J.-F., Eds.; Advances in
  Polymer Science 12; Springer Berlin Heidelberg: Berlin, Heidelberg, 2009;
  Vol.~58; pp 1--71\relax
\mciteBstWouldAddEndPuncttrue
\mciteSetBstMidEndSepPunct{\mcitedefaultmidpunct}
{\mcitedefaultendpunct}{\mcitedefaultseppunct}\relax
\EndOfBibitem
\end{mcitethebibliography}
\end{document}